\documentclass[final,5p,times,twocolumn]{elsarticle}
\usepackage{amssymb}
\biboptions{numbers,sort&compress}

 \usepackage{amsmath}
\usepackage[utf8x]{inputenc}
\usepackage[T1]{fontenc}
\usepackage{amsfonts} 
\usepackage{amsfonts} 
\usepackage{mathrsfs}
\usepackage{braket}
\usepackage{graphicx}

\def\beq{\begin{equation}}
\def\eeq{\end{equation}}
\def\bsp{\begin{split}}
\def\esp{\end{split}}
\def\bea{\begin{eqnarray}}
\def\eea{\end{eqnarray}}
\def\ba{\begin{array}}
\def\ea{\end{array}}

\def\sg{\sigma}

\def\lb{\left(}
\def\rb{\right)}

\def\l.{\left.}
\def\r.{\right.}

\def\x{\times}

\journal{Physica B}
\begin{document}

\begin{frontmatter}

\title{Theoretical Studies of Dimeric Exchange-Coupled Single-Molecule Magnets}

\author[aff1,aff2]{S. A. Owerre}
\ead{solomon.akaraka.owerre@umontreal.ca}
\address[aff1]{D\'epartement de physique,
Universit\'e de Montr\'eal,
Montr\'eal, 
Qu\'ebec H3C 3J7, Canada.}


\begin{abstract}
We study a molecular dimer model that takes higher order interaction term as well as antisymmetric term into account. We mainly focus on the theoretical quantum properties that are of interest. We numerically diagonalize the system and investigate different phenomena which include oscillations of ground state energy splitting, magnetization steps, Schottky anomaly, torque oscillations, and magnetocaloric effects. Indeed, we show that these interactions can exotically influence these phenomena, which should be of interest. 
\end{abstract}

\begin{keyword}
  Molecular magnets\sep Tunneling \sep Antiferromagnetism\sep Magnetocaloric effects.
\end{keyword}
\end{frontmatter}

\section{Introduction} 
The study of molecular magnets has captivated the attention of many researchers in recent years. It has proven be a ubiquitous research area in physics, with applications ranging from quantum information processing   \cite{AB}, molecular spintronics   \cite{AC, AD} to molecule-based magnetic coolers   \cite{AE, AF, affr} (magnetocaloric effects). Indeed, these molecular magnets are composed of very large spins with large anisotropy, which makes their application to technology more feasible. The large anisotropy in these magnets is responsible for an energy barrier  between two degenerate spin states,  and at very low temperature macroscopic quantum tunneling is, indeed, possible between the lowest states on either sides of the potential minima   \cite{AG, AH, AI, loss1}. These magnets garnered so much attention of researchers when it was predicted serendipitously, that for integer spins tunneling is allowed, while for half-odd integer spin tunneling is completely suppressed at zero (external)  magnetic field   \cite{loss1}.  The vanishing of tunneling for half-odd integer spins is understood as a consequence of destructive interference between tunneling paths, which is directly related to Kramers' degeneracy   \cite{kram}, due to the time reversal invariance of the Hamiltonian. In the presence of a magnetic field applied along the spin hard axis direction, Garg   \cite{anu2} showed that the tunneling splitting does not vanish for half-odd integer spins. In this case,  different tunneling paths along the hard axis direction accumulate phases which add constructively and leads to an oscillation of the energy splitting as a function of the magnetic field. The tunneling splitting only vanishes at some critical values of the field, which, of course,  is not related to  Kramers' degeneracy due to the presence of a magnetic field. These serendipitous theoretical predictions were subsequently observed experimentally in Fe$_8$ molecular cluster   \cite{wern}.  An exposition of these enthralling scientific results  and other potential technological applications has been reviewed recently   \cite{sol, sol1, sol2}. However, the research in this field is far from over; there are still much interesting phenomena to be investigated in these magnets. 

In the past few decades, the research in this field has mainly focused on single molecule magnets. In many cases of physical interest, however,  interactions between two giant spins are not negligible. At the lowest order, the simplest interaction is of the symmetric bilinear Heisenberg form, which can be  either ferromagnetic or antiferromagnetic depending on the sign of the exchange constant.  This interaction is manifested explicitly in some physical systems such as molecular wheels Mn$_{12}$, Na:Fe$_6$, Cs:Fe$_8$  \cite{wal,mula, had, cin}, and the molecular dimer [Mn$_{4}$]$_2$  \cite{da,aff}. These systems have been studied theoretically via the semiclassical spin coherent state path integral formalism   \cite{bar, ml2001}, in which an instanton trajectory mediates tunneling of N\'eel vectors when the barrier height is larger than the ground state energy. Experimentally, the lowest order  giant spin model  approximation is inadequate in explicating experimental observations  \cite{cin, carr, da1, aff}. The predicted splittings of macroscopic quantum tunneling resonances turn out to be less than the inhomogeneous linewidths of the hysteresis steps  \cite{da1, aff}, which proves the inefficacy of the lowest order approximation. Thus, in order to corroborate experimental observations, it is expedient to include other exotic terms in the Hamiltonian.  These additional terms  possess markedly different behaviours which can  change the quantum nature the system  \cite{wern2}. For the case of  dimeric molecular magnet, these terms involve the antisymmetric Dzyaloshinskii-Moriya (DM) exchange interaction, biquadratic exchange interaction, etc. These terms, however, pose difficult analytical solutions. In fact, it is infeasible to find an instanton trajectory that mediates the tunneling of N\'eel vectors.

In this paper we give an explication of the quantum behaviour of a dimer model which takes these interactions into account, in the presence of an easy axis anisotropy and a magnetic field applied along the longitudinal and transverse directions. In order to exemplify our approach, we will specialize in the case of two equal spins $s_A=s_B=9/2$, though our analysis can be extended to lower or higher spin systems.  Molecular dimers with $s_A=s_B=9/2$ include [Mn$_{4}$]$_2$  \cite{da,aff}, Mn$_{4}$Cl and Mn$_{4}$Me  \cite{evan}. It is worth pointing out that we do not intend to corroborate any experimental results that have been demonstrated in these dimers;  we  will, however, be interested in different properties that can be influenced by the biquadratic and the antisymmetric exchange terms, which in principle should be of interest.   We will investigate different quantum and thermodynamic properties in these systems via an exact numerical diagonalization technique. We also explore  thoroughly the parameter space of this model. 

The model Hamiltonian  will be presented in Sec.(II); while Sec.(III) gives a synopsis of the analytical solution of the N\'eel vector tunneling.  The effects of the interactions,  the longitudinal and transverse magnetic field on the thermodynamic quantities will be presented in Sec.(IV) and Sec.(V) via an exact numerical diagonalization. Indeed, we observe the plateaus in the magnetization and the Schottky anomaly in the magnetic specific heat. When a molecular magnet is placed in a magnetic field, the magnetic moment or magnetization experiences a torque and the magnetic anisotropy can be measured by torque magnetometry  \cite{con} which gives rise to quantum oscillation. This phenomenon will be theoretically investigated in Sec.(VI). The variation of an applied field on a magnetic material can remarkably change its temperature in an adiabatic process, as well as its entropies by absorbing or releasing heat in an isothermal process. This is the basic principle of the magnetocaloric effect  \cite{von} (MCE), which will be presented theoretically in Sec.(VII).
\section{Model} 
In this section,  we will  present a dimeric molecular magnet of equal spins with three interaction term --- an easy axis anisotropy, and a magnetic field. The Hamiltonian of this system has the simple form:\begin{align}
\mathcal{\hat{H}}&= H_{0} + H_{\text {BQ}} + H_{\text{AS}} + H_{\text{AN}}+ H_{Z},
\label{1}
\end{align}
where
\begin{align}
&H_{0}=J\hat{\bold{S}}_A \cdot \hat{\bold{S}}_B,
\label{hes}
\end{align}
is the isotropic bilinear interaction term between two giant spins of the molecular dimer, with $J>0$ being antiferromagnetic coupling  and $J<0$ being ferromagnetic coulping.
The biquadratic term is given by
\begin{align}
H_{\text {BQ}}= \mathcal{J}\lb\hat{\bold{S}}_A \cdot \hat{\bold{S}}_B\rb^2.
\label{bi}
\end{align}
In most dimeric molecules, it is expedient to include this term as a correction to the bilinear exchange term in Eq.\eqref{hes} with $\mathcal{J} \ll J$.  The effects of this term has been recently reported   \cite{fur} in a magnetic diluted compound KMn$_{0.1}$Zn$_{0.9}$F$_3$. The antisymmetric Dzyaloshinskii-Moriya exchange interaction has the usual form:\begin{align}
H_{\text{AS}}= \bold{D}\cdot\lb\hat{\bold{S}}_A \times\hat{\bold{S}}_B\rb.
\label{dm}
\end{align}
This kind of interaction was first proposed by Dzyaloshinskii   \cite{U} and subsequently derived by Moriya   \cite{V}. It is responsible for the anticrossing mechanism observed in many molecular magnets    \cite{cin}.  It has been demonstrated that this antisymmetric interaction induces transition between different spin multiplets in single molecule magnets    \cite{wern2}. Its existence, however, depends on the symmetry of the system. It signifies a broken inversion symmetry in molecular magnets.  In many systems, it is customary to restrict the interaction along the $z$ axis, {\it i.e}, $\bold{D}= D\bold{z}$, as other components do not lead to anticrossing  between energy levels. The anisotropy term is taken to be of the easy axis form:
\begin{align}
H_{\text{AN}}=
  - \mathcal{K}(\hat{S}_{A,z}^{2}+\hat{S}_{B,z}^{2}); \quad \mathcal{K}>0.
  \label{ani}
\end{align}
This term is responsible for an energy barrier which separates two eigenstates of the spin system. We have neglected a fourth-order axial operator for simplicity. The Zeeman term can be written as
\begin{align}
H_{Z}=-g\mu_B H_{||} (\hat{S}_{A,z}+\hat{S} _{B,z}) -g\mu_B H_{\perp}\cos\phi(\hat{S}_{A,x}+\hat{S} _{B,x}),
\label{zee}
\end{align}
where $\phi$ is the angle between the magnetic field and total $x$ axis; $g\approx 2$ is the spin $g$-factor, and $\mu_B$ is the Bohr magneton.  Without loss of generality we have taken the $y$ component in Eq.\eqref{zee} to be zero.

The model defined in Eq.\eqref{1} is quite general. R. Tiron {\it et al} \cite{da} have modelled the molecular dimer  [Mn$_{4}$]$_2$ with $\mathcal{H}=H_0 + H_{\text{AN}} + H_Z$. They determined that this dimer possesses a strong anisotropy  with $\mathcal{K}=0.77K$ and $J=0.13 K$. In the subsequent sections, we will produce most of our figures by keep these values fixed. However, most molecular magnets have dominant bilinear interaction, $J>\mathcal K$. We will also work in this limit. For the additional terms, we will assume them to be small, and see the effects they introduce into the systems. 

\section{N\'eel  Tunneling}
One of the enthralling properties of antiferromagnetic molecular magnets is the possibility of N\'eel  tunneling    \cite{bar, ml2001}. This is usually investigated via the semiclassical approach. In this formalism the classical energy corresponding to Eq.\eqref{1} can be written as
\begin{align}
U(\bold{\hat n})&= J\hat{\bold{n}}_A\cdot \hat{\bold{n}}_B +\mathcal{J}\lb \hat{\bold{n}}_A\cdot \hat{\bold{n}}_B\rb^2+\bold{D}\cdot\lb\hat{\bold{n}}_A \times\hat{\bold{n}}_B\rb\nonumber\\&- \mathcal{K}(\hat{n}_{A,z}^{2}+\hat{n}_{B,z}^{2})-g\mu_B H_{||} (\hat{n}_{A,z}+\hat{n} _{B,z}) \nonumber\\&-g\mu_B H_{\perp}\cos\phi(\hat{n}_{A,x}+\hat{n} _{B,x}),
\label{neel}
\end{align} 
where $\hat{\bold{n}}$ is a unit vector that parameterizes the two-sphere. The imaginary time propagator is of the form:
\bea
\braket{\hat{\bold n}_f|e^{-\beta \hat{\mathcal{H}}}|\hat{\bold n}_i}= \int  \mathcal{D}\hat {\bold n}\thinspace e^{-S_E},
\eea

where the Euclidean action is given by
\begin{equation}
S_E[\bold{\hat n}] = isS_{{WZ}}+ \int d\tau U(\bold{\hat n(\tau)});
\label{act1}
\end{equation}
with
\begin{align}
S_{WZ} =is\sum_{j}\int d\tau \frac{1}{1+n_{j,z}}(n_{j,x}\partial_\tau n_{j,y}-n_{j,y}\partial_\tau n_{j,x}),
\label{3.13a}
\end{align}
and $j=A,B$. The coordinate dependent form of Eq.\eqref{3.13a} can be easily recovered by expressing the unit vectors using spherical coordinate parameterization. 
For a given potential energy of the system, $U(\bold{\hat n(\tau)})$, one is  interested in finding the solution for $\bold{\hat n(\tau)}$  that interpolates between two degenerate minima or metastable states of the potential energy. This leads to the so-called ``macroscopic quantum tunneling'', which is mediated by an instanton trajectory. In the present problem, however, it is cumbersome to find  an exact instanton trajectory that mediates tunneling. In the limit $\mathcal{J}=D=H_{\parallel}=0$; this problem has been explicitly solved analytically by integrating out the fluctuations around the N\'eel vector. Quantum tunneling, in this case, lifts the degeneracy of the ground states and produces an energy splitting, which oscillates as a function of the transverse magnetic field     \cite{bar, ml2001}.
\section{Numerical Diagonalization}
 The numerical diagonalization of this system relies mainly on the proper choice of basis in the Hilbert space, from which the matrix can be constructed.   The Hilbert space $\mathscr{H}$ of the total Hamiltonian in Eq.\eqref{1} is the tensor product of the two spaces: $\mathscr{H}=\mathscr{H}_A \otimes \mathscr{H}_B$, with dim$(\mathscr{H})$= $(2s_A+1)\otimes (2s_B+1)$. In this product space, a convenient  basis can be written as:  \bea
  \ket{s_A, \sigma_A}\otimes\ket{s_B, \sigma_B} \equiv\ket{\sigma_A,\sigma_B},
  \label{bas}
  \eea
\begin{figure}[ht]
\centering
\includegraphics[width=3.5in]{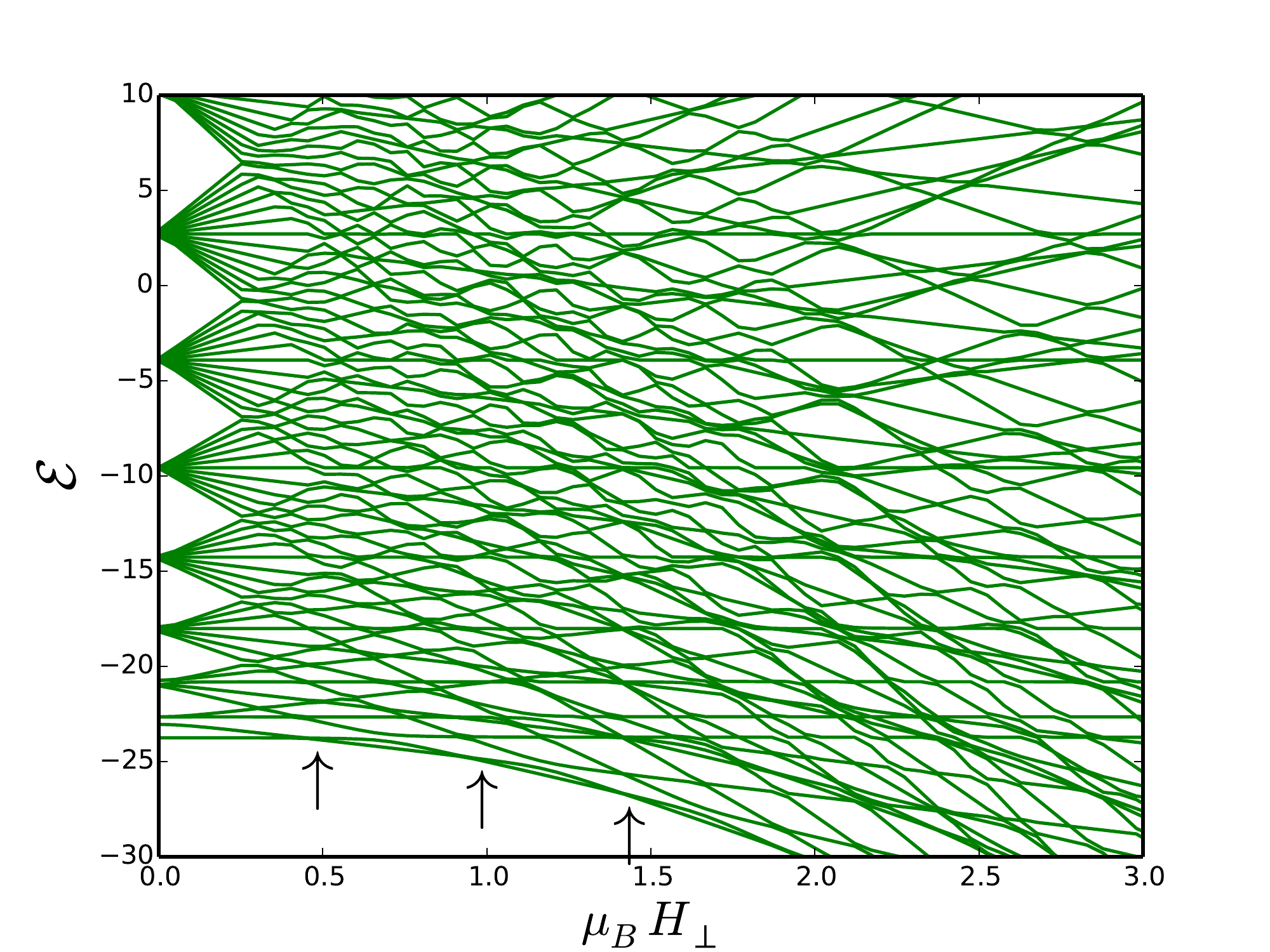}
\includegraphics[width=3.5in]{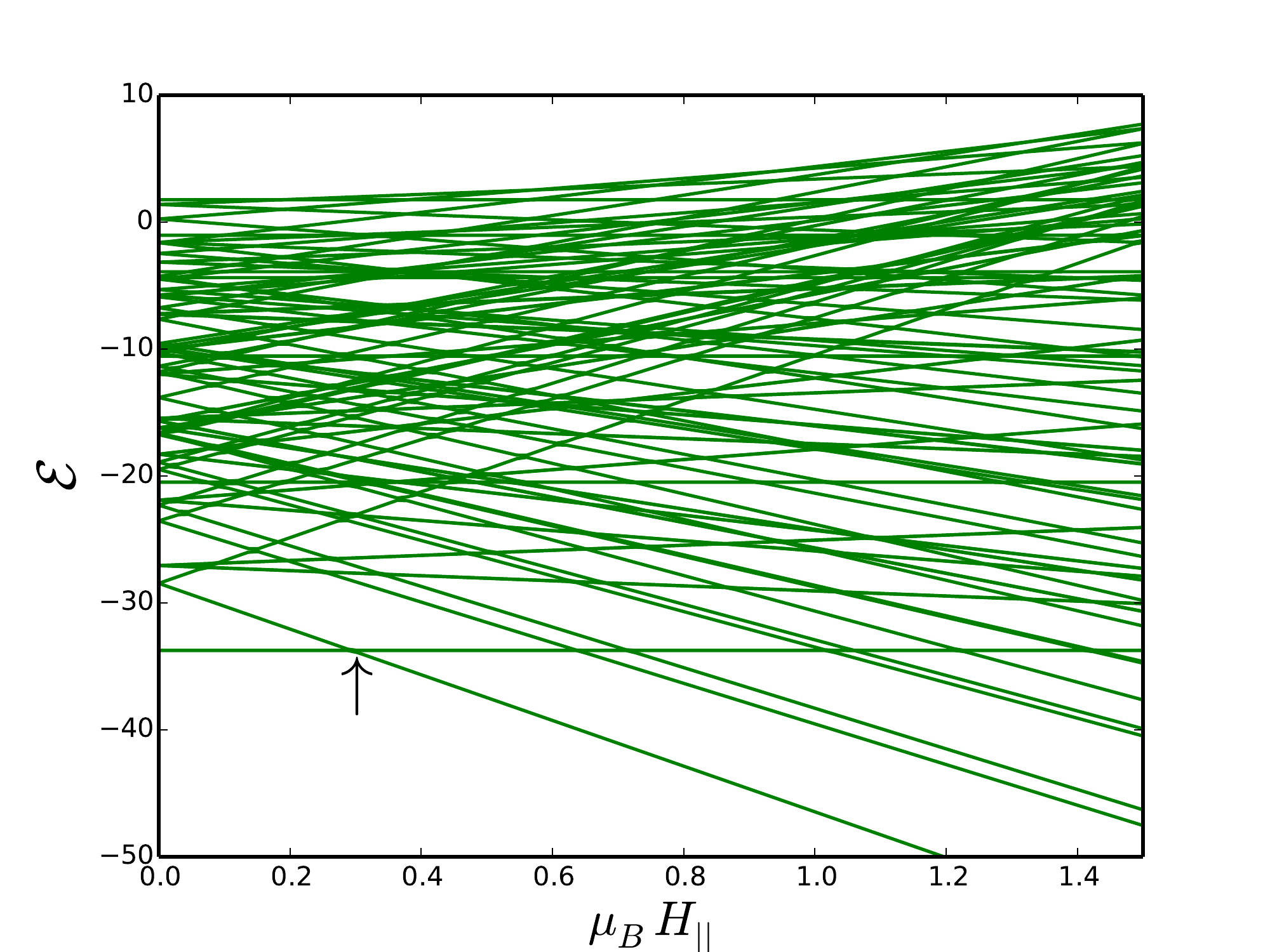}
\caption{The $100$ nonzero eigenvalues, $\mathcal{E}(\text{Kelvins})$ of Eq.\eqref{1} plotted against the magnetic field, $H(\text{Tesla})$. Top: the transverse field. Bottom: the longitudinal field at $\phi=0$. Top figure parameters: $\mathcal{K}=0.02[K], ~J=0.95[K], ~D= 3\times 10^{-3}[K],~ \mathcal{J}=2.3\times 10^{-4}K, ~H_{||}=0$. Bottom figure parameters: $\mathcal{K}=0.77[K], ~J=0.13[K], ~D= 3\times 10^{-3}K, ~\mathcal{J}=2.3\times 10^{-4}[K], H_{\perp}=0$. The arrows indicate the crossings of the ground state and the first excited state.}
\label{eig}
\end{figure}
\begin{figure}[ht]
\centering
\includegraphics[width=3.5in]{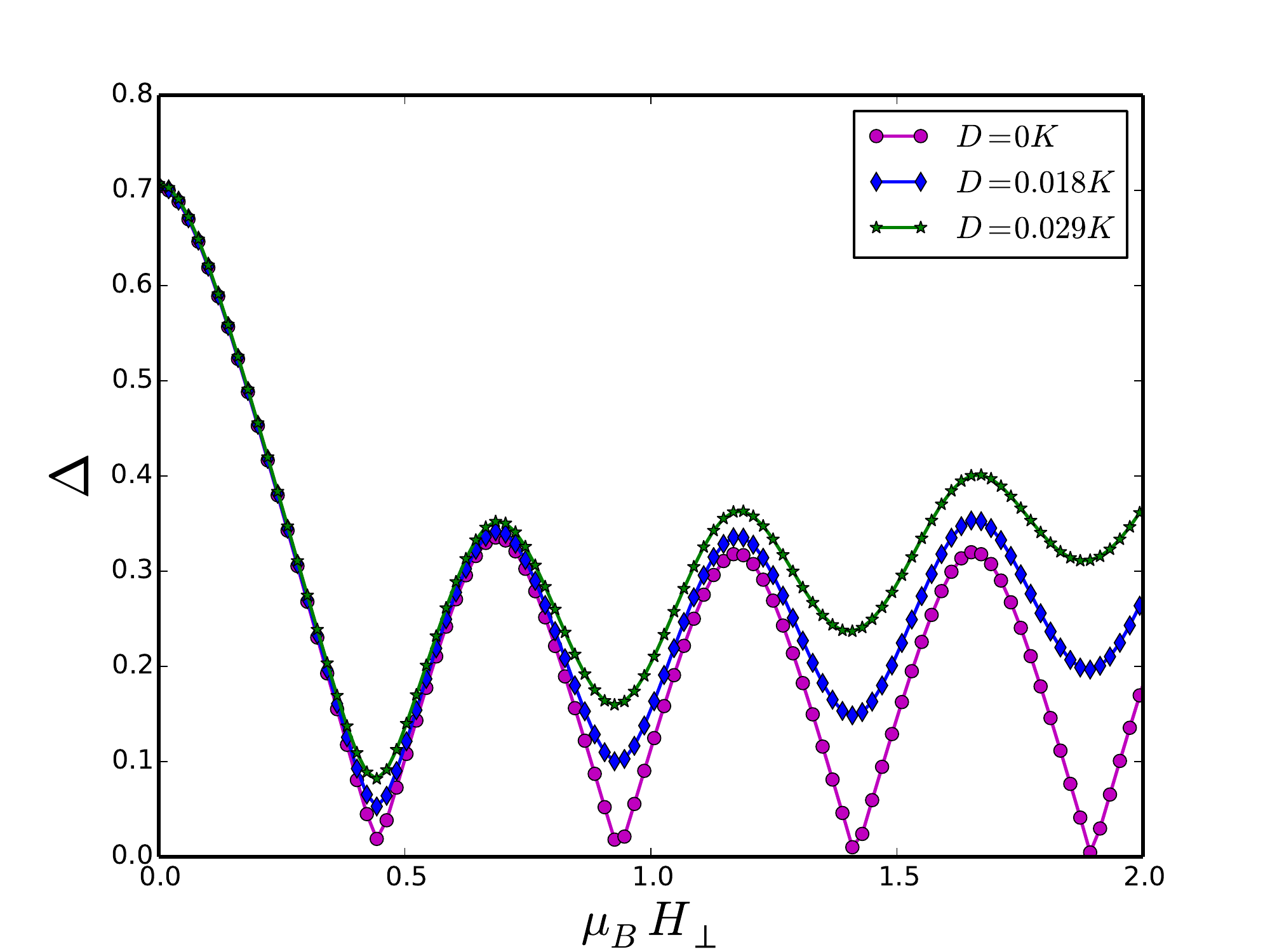}
\includegraphics[width=3.5in]{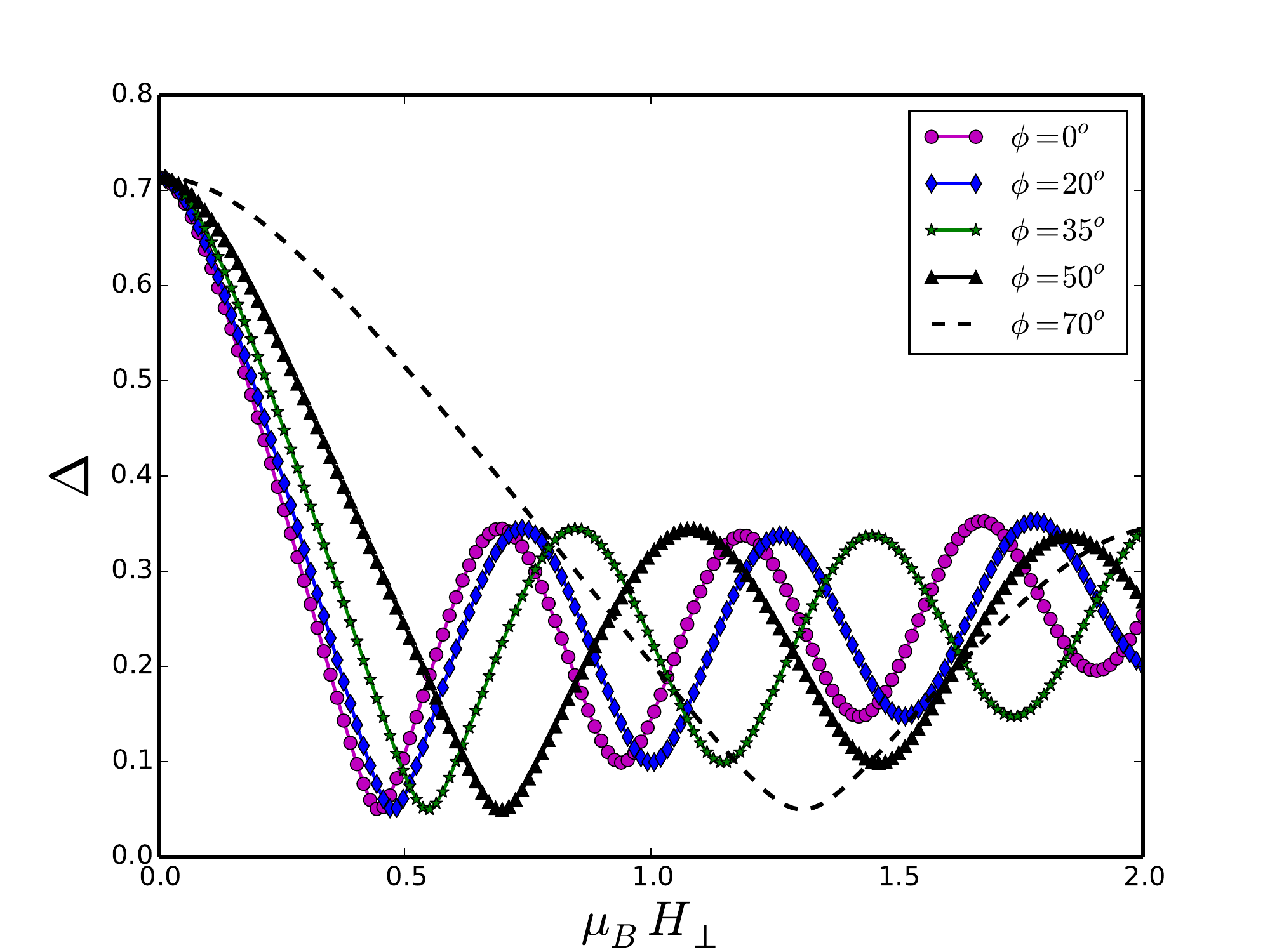}
\caption{The ground state energy splitting $\Delta([K])$ plotted against the transverse field, $H_\perp(\text{Tesla})$ for several values of $D$ (Top) and several values of $\phi$ (Bottom). Top figure parameters: $\mathcal{K}=0.02[K],~ J=0.95[K],  \mathcal{J}=2.3\times 10^{-4}[K], \phi=0, H_{||}=0$. Bottom figure parameters: $\mathcal{K}=0.02[K],~ J=0.95K, ~D= 0.018,~ \mathcal{J}=2.3\times 10^{-4}[K],~ H_{||}=0$.}
\label{split}
\end{figure}

where $\sigma_A=-s_A,-s_A+1,\cdots, s_A$ and $\sigma_B=-s_B,-s_B+1,\cdots, s_B$. Specializing in the case of equal spins $s_A=s_B= 9/2$, the matrix elements of Eq.\eqref{1} correspond  to a $100\x 100$ sparse matrix, which can be constructed either from Eq.\eqref{bas} or by using the general spin wavefunction of the system    \cite{solm}. In the absence of the transverse terms (raising and lowering operators) in Eq.\eqref{1}, the energy levels are given by the diagonal matrix elements:
\begin{align}
&\mathcal{E}_{\sg_A,\sg_B}=-\mathcal{K}(\sg_A^2+\sg_B^2)-g\mu_BH_{||}(\sg_A+\sg_B)+J\sg_A\sg_B\nonumber\\&+\mathcal{J}\bigg[ (\sg_A\sg_B)^2 + \frac{(s+\sg_A)(s-\sg_A+1)(s-\sg_B)(s+\sg_B+1)}{4}\nonumber\\&+ \frac{(s-\sg_A)(s+\sg_A+1)(s+\sg_B)(s-\sg_B+1)}{4}\bigg].
\end{align}
The corresponding eigenstates are given by Eq.\eqref{bas}. At zero magnetic field the lowest energy states are the two N\'eel states $\ket{s_A,-s_B}$ and $\ket{-s_A,s_B}$ assuming $J> \mathcal{J}> 0$.   With a nonzero magnetic field, states with  $\sg_A+\sg_B >0$ decrease while those with $\sg_A+\sg_B <0 $ increase. Thus, some of the levels cross each other at some values of the magnetic field. The resonance condition for doubly degeneracy of these levels is given by
\bea
\mathcal{E}_{\sg_A,\sg_B}=\mathcal{E}_{\sg_A^{\prime},\sg_B^{\prime}},
\label{reso}
\eea
which determines the values of the longitudinal magnetic field $H_{||}$ at which energy levels cross each other. These degeneracies are lifted (avoided energy crossing) via macroscopic quantum 
tunneling, which is mediated  by the raising and the lowering operators stemming from  Eqs.\eqref{hes}, \eqref{bi}, \eqref{dm} and \eqref{zee}, and the corresponding eigenstates become linear superpositions of the degenerate states.  Indeed, the transverse field induces transition between levels with $\Delta\sg_A=\pm 1$, $\Delta\sg_B=0$ and vice versa; the transverse terms emanating from Eqs.\eqref{hes} and \eqref{dm} induce transition between levels with $\Delta\sg_A=\pm 1$, $\Delta\sg_B=\pm 1$, while the transverse terms from Eq.\eqref{bi} induce transition between levels with $\Delta\sg_A=\pm 1$, $\Delta\sg_B=\pm 1$ and $\Delta\sg_A=\pm 2$, $\Delta\sg_B=\pm 2$.

 The exact numerical diagonalization of Eq.\eqref{1} leads to the energy splitting shown in Fig.\eqref{eig}. With dominant anisotropy, several levels cross each other, while dominant interaction term leads to many avoided level crossings.  Quantum tunneling is evident from these figures with the ground state $\ket{\mathcal{E}_0}=\frac{1}{\sqrt{2}}\lb\ket{-\frac{9}{2}, \frac{9}{2}}- \ket{\frac{9}{2}, -\frac{9}{2}}\rb$ crossing the first excited state $\ket{\mathcal{E}_1}=\frac{1}{\sqrt{2}}\lb\ket{-\frac{9}{2}, \frac{9}{2}}+ \ket{\frac{9}{2}, -\frac{9}{2}}\rb$ at several intervals of the transverse field as indicated by the arrows. These crossings lead to an oscillation of the ground state energy splitting as shown in Fig.\eqref{split}. The increase of the antisymmetric interaction leads to an increase in the amplitude of the oscillation,  whereas an increase in the angle decreases the number of oscillations.
  \section{Magnetization and magnetic specific heat capacity}
 In this section, we will study the thermodynamics of our system.  To compute any thermodynamic quantity, it is customarily expedient to first calculate the partition function of the system. The canonical partition function is given by
 \bea
 Z= Tr(e^{-\beta\mathcal{H}}) =\sum_{i=1}^{(2s+1)^2}e^{-\beta\mathcal{E}_i},
 \label{pat}
 \eea
 \begin{figure}[ht]
\centering
\includegraphics[width=3.5in]{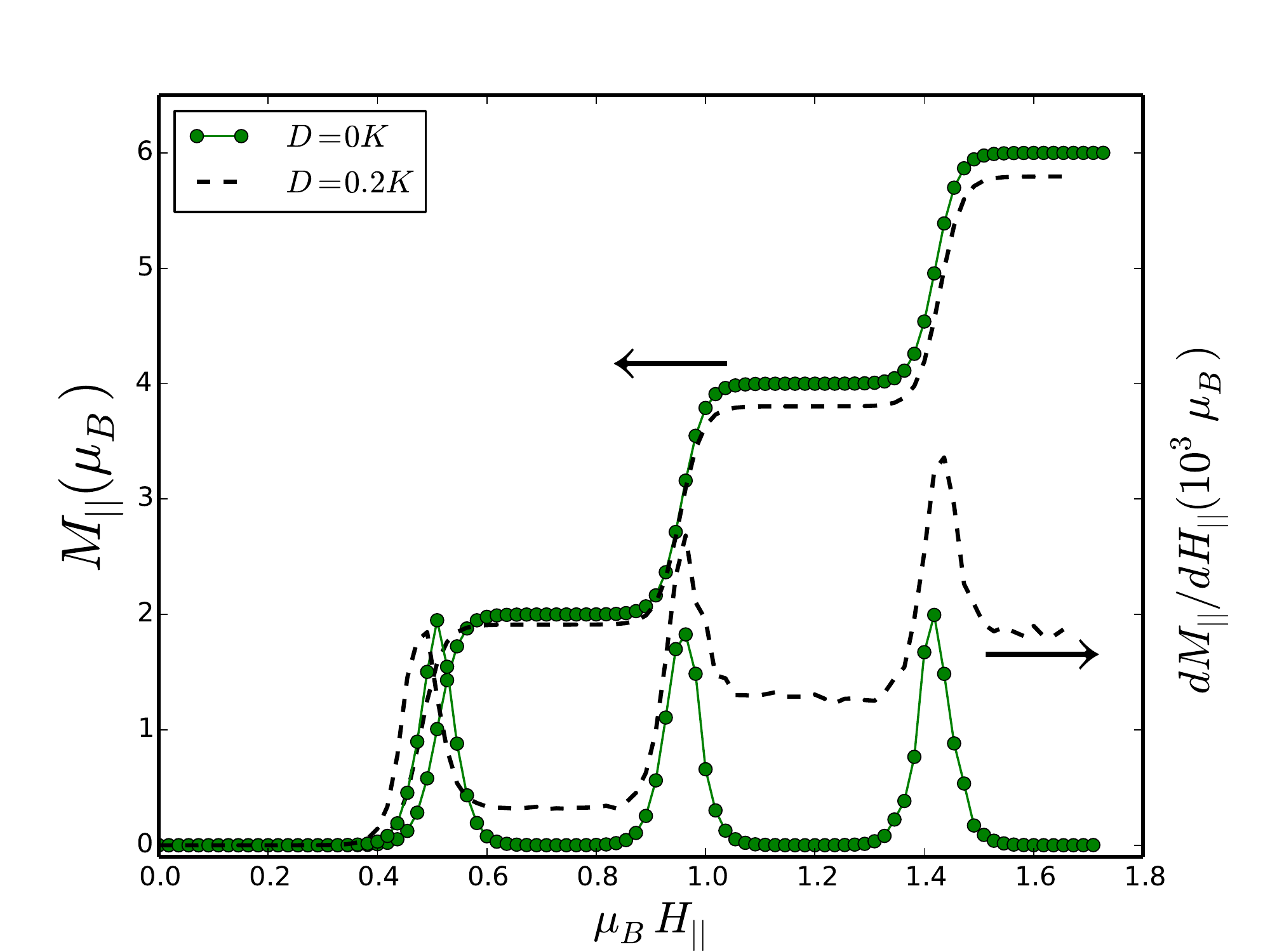}
\includegraphics[width=3.5in]{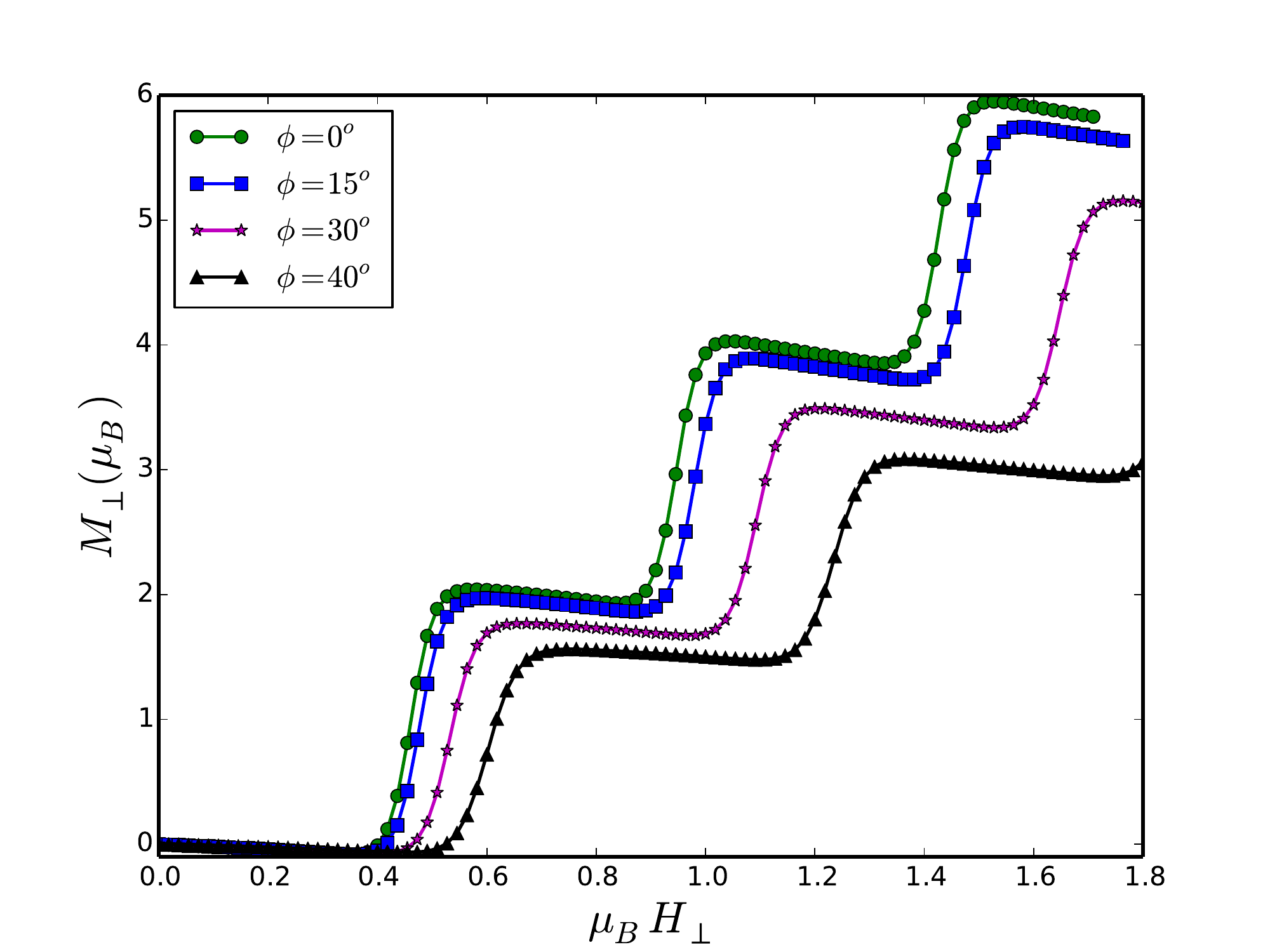}
\caption{Top figure. The plot of the longitudinal magnetization  and the corresponding differential susceptibility in units of $\mu_B$ against the transverse field, $H_\perp(\text{Tesla})$,  with $\mathcal{K}=0.01[K], ~ J=0.95[K], ~\mathcal{J}=3.3\times 10^{-5}[K], \phi=0, ~T=0.04[K],~ H_{\perp}=0$. Bottom figure. The plot of the  transverse magnetization in in units of $\mu_B$ against the transverse field, $H_\perp(Tesla)$,  with several values of the azimuthal angle $\phi$, with the same parameters except $D= 0.003[K]$ and $H_{||}=0$.}
\label{mag}
\end{figure}

where  $\beta= T^{-1}$ is the inverse temperature of the system and $\mathcal{E}_i$ are the eigenvalues. Now, the thermodynamic quantities of interest can conveniently be derived from Eq.\eqref{pat}. The total magnetization of our system is given by
\bea
M_{\alpha}= \frac{1}{Z}\sum_{i=1}^{(2s+1)^2}\braket{\mathcal{E}_i|S_{\alpha}|\mathcal{E}_i}e^{-\beta\mathcal{E}_i}=\frac{1}{\beta}\frac{\partial \ln Z}{\partial H_{\alpha}},
\eea
where $\mathcal{E}_i$ and  $ \ket{\mathcal{E}_i}$ are the eigenvalues and eigenfunctions of the Hamiltonian in Eq.\eqref{1} respectively; $S_{\alpha}$ are the components of  the total spin  and $H_\alpha$ are the components of the magnetic field with $\alpha=x,y,z$. Fig.\eqref{mag} shows the longitudinal magnetization with zero and nonzero DM interaction. Indeed, the magnetization develops step-like increase at low-temperature,  while the differential susceptibility shows sharp peaks at each step of the magnetization curve. The step-like increase in the magnetization and corresponding sharp peaks in the susceptibility occur at the values of $H_{||}$ that solves Eq.\eqref{reso}. A small nonzero  DM interaction decreases the steps and increases the peaks of the susceptibility.   A considerable increase in the DM and the biquadratic interactions obliterates the magnetization steps, thus exterminates the sharp peaks in the susceptibility. The transverse susceptibility as well possesses plateaus which diminish as the azimuthal angle $\phi$ increases; however, there are no evident peaks on the susceptibility curves as a consequence  of  Eq.\eqref{reso}.

Next, we will consider the specific heat capacity of the system. The specific heat of a material is a ubiquitous  quantity in condensed matter physics. It provides significant information on various system excitations. In experimental measurements, the specific heat has  lattice, electronic, and magnetic  contributions. One measures a particular contribution by measuring the total specific heat of the system, and subtracting the unwanted contributions. The magnetic contribution is given by
\bea
C_{mag}= \frac{\partial \braket{\mathcal E}}{\partial T},
\eea
where the average energy $\braket{\mathcal E}$ is given by
\bea
\braket{\mathcal E}=\frac{1}{Z}\sum_{i=1}^{(2s+1)^2}\mathcal{E}_ie^{-\beta\mathcal{E}_i}.
\eea
\begin{figure}
\centering
\includegraphics[width=3.5in]{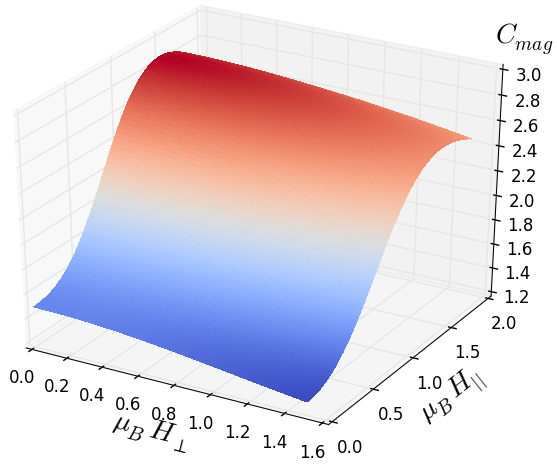}
\caption{Three dimensional plot of the magnetic specific heat, $C_{mag}(J/kg)$ as a function of the longitudinal field, $H_\parallel(\text{Tesla})$ and the transverse field  $H_\perp(\text{Tesla})$ for $\mathcal{K}=0.77[K],~ J=0.13[K], ~D=0.03, ~\mathcal{J}=3.3\times 10^{-4}[K], ~\phi=0, ~T=6.5[K]$.}
\label{cmag}
\end{figure}
\begin{figure}
\centering
\includegraphics[width=3.5in]{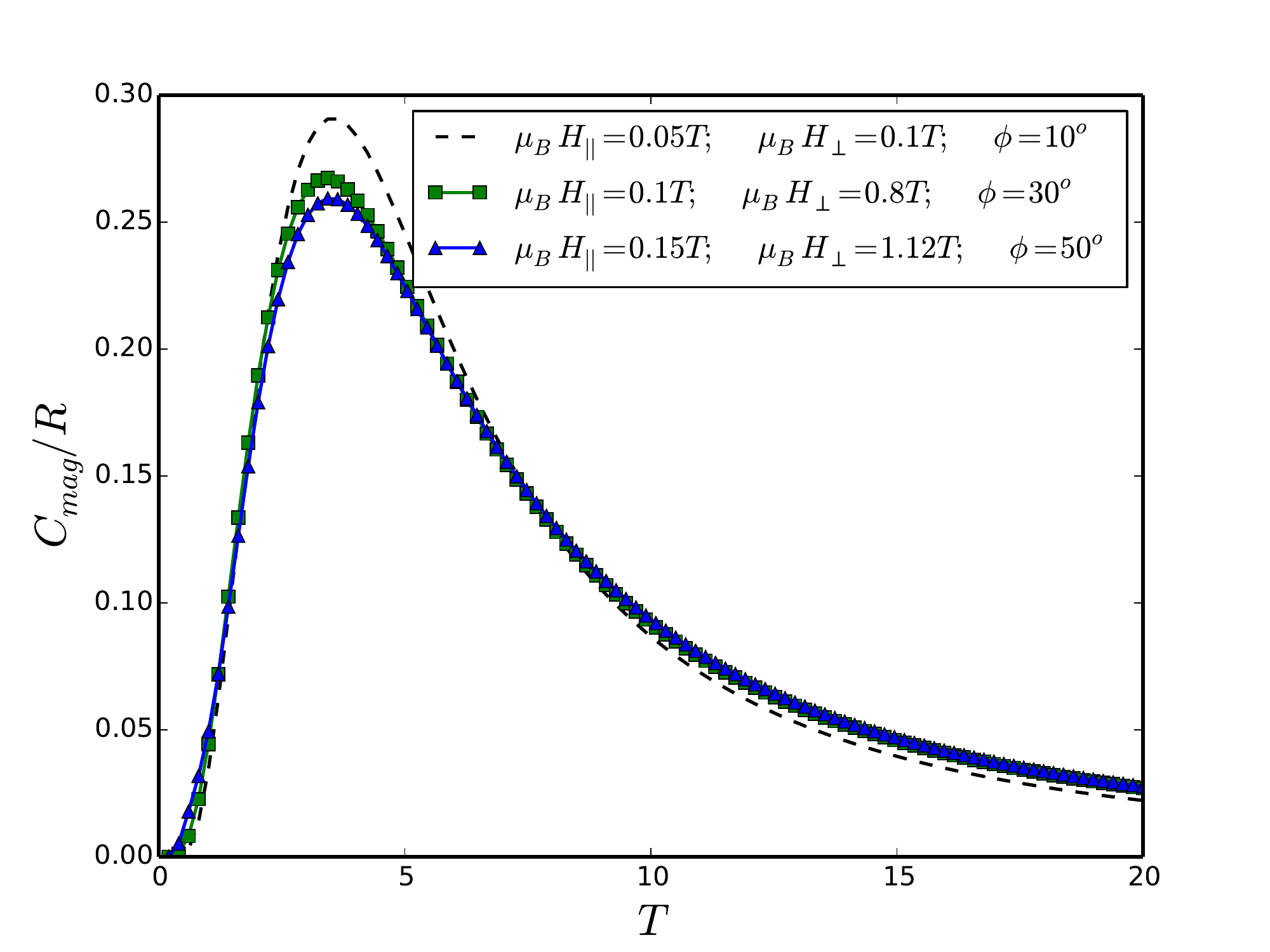}
\caption{Magnetic specific heat, $C_{mag}(J/kg)$, vs.Temperature, $T([K])$ for $\mathcal{K}=0.77[K],~ J=0.13[K],~ D=0.11[K],~ \mathcal{J}=2.3\times 10^{-4}[K]$. We have normalized the magnetic specific by the gas constant $R$.}
\label{cmag1}
\end{figure}
We noticed that the magnetic specific heat shows a decrease along the transverse field direction and increases along the longitudinal field direction as shown in Fig.\eqref{cmag}. At the magnetization steps, the magnetic specific heat oscillates as a function of the magnetic field.  
The temperature dependence of the magnetic specific heat in Fig.\eqref{cmag1} exhibits the usual behaviour of Schottky anomaly  \cite{gop} as a consequence of the $(2s+1)^2$ discrete energy level which are split in a magnetic field  as shown in Fig.\eqref{1}. The specific heat goes to zero at low and high temperatures due to low and high populations of the discrete levels respectively, but it exhibits a sharp peak at the intermedate  $T_m\sim 5K$, which is related to the energy level splitting. The decrease of the magnetization steps as the DM interaction increases moderately also manifest in the specific heat (not shown).
\section{Quantum oscillation}
\begin{figure}
\centering
\includegraphics[width=2in]{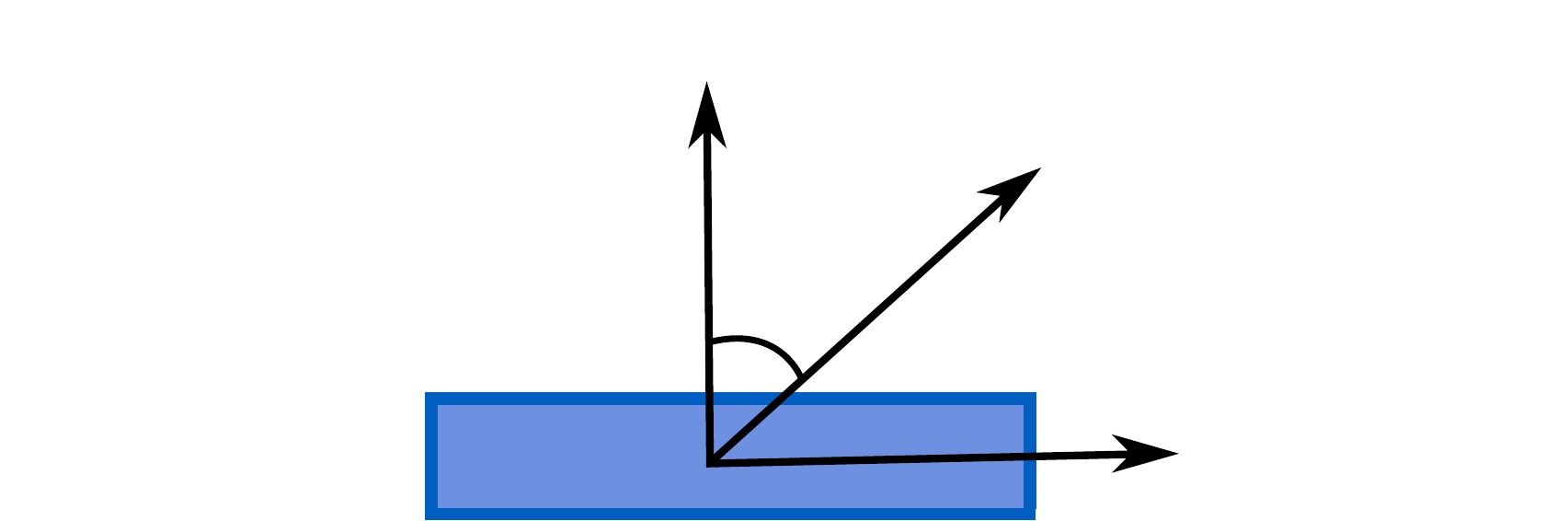}
\caption{ Schematic arrangement of a sample in a magnetic field on the $xz$ plane. The angle $\theta$ is between the magnetic field and the easy $z$-axis.} 
\label{tor}
\end{figure}

One of the experimental techniques to measure the anisotropy of magnetic susceptibility of molecular magnets is the torque magnetometry  \cite{con}. This technique is based on the basic principle that when a magnetic sample is placed in a uniform magnetic field, the magnetic moment or magnetization experiences a force ; see Fig.\eqref{tor}. Thus, a torque is generated, which is given by
\bea
\boldsymbol{\tau}= \bold{M}\times\bold{B}.
\label{to}
\eea
\begin{figure}[ht]
\centering
\includegraphics[width=3.5in]{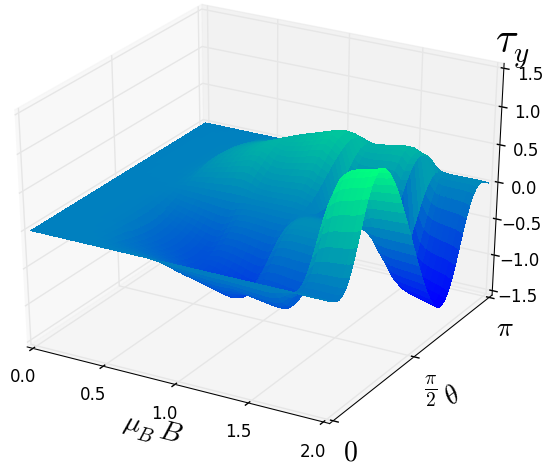}
\caption{Three dimensional plot of the magnetic torque in units of $\mu_B$ as a function of the angle $\theta$ and the magnetic field for $\mathcal{K}=0.01[K], ~J=0.85[K], ~D= 1.2\times 10^{-4}[K],~ \mathcal{J}=3.3\times 10^{-4}K,~T=0.05[K]$.}
\label{torq1}
\end{figure}
In a typical experimental setup, it is customary to apply the rotation of the torquemeter  parallel to the $y$-axis, thus the only observable of interest is the $y$-component of the torque. The new Hamiltonian for our model under the action of a magnetic field in the $x$-$z$ plane at an angle $\theta$ has the form:
\begin{align}
\mathcal{\hat{H}}&= H_{0} + H_{\text {BQ}} + H_{\text{AS}} + H_{\text{AN}}+ g\mu_B B(S_z\cos\theta + S_x\sin\theta),
\label{11}
\end{align}
where $B=\sqrt{B_z^2+B_x^2}$, $S_z=S_{z,A}+S_{z,B}$, and $S_x=S_{x,A}+S_{x,B}$. It is easy to discern the relationship between Eq.\eqref{1} and  Eq.\eqref{11}. 
The $y$ component of the torque can be computed directly from Eq.\eqref{to} or by the simple formula  \cite{con}:
\begin{align}
\tau_y = -\lb\frac{\partial \braket{\mathcal{H}}}{\partial\theta}\rb_B=g\mu_B B\lb \braket{S_z}\sin\theta-\braket{S_x}\cos\theta\rb.
\end{align}
The average values correspond to the magnetization in the $x$ and $z$ directions,  which  can now be computed using the $100$ eigenfunctions and eigenvalues of the Hamiltonian in Eq.\eqref{11}. As shown in Fig.\eqref{torq1}, the torque oscillates as a function of the angle $\theta$ at $T=0.05K$, and vanishes when the magnetic field is parallel to the easy axis or the hard axis, {\it i.e}, $\theta=0$ or $\theta=\pi/2$ respectively, for all values of the magnetic field. Along the magnetic field it exhibits step-like structure at the crossing fields $B_c$ with a dominant bilinear interaction. A moderate increase in the DM interaction decreases both the oscillations and the step-like structures.

\section{Magnetocaloric effect}
Magnetocaloric effect (MCE) is the principle that governed magnetic refrigeration  \cite{von, AE, affr}. A description of this effect is achieved by a change in the isothermal entropy $\Delta S_m$ or an adiabatic temperature change $\Delta T_{ad}$ as a result of change in the magnetic field $\Delta B$.  Magnetic materials that maximize $\Delta S_m$ are deemed to be more effective for magnetic refrigerants.    The isothermal entropy change can as well be maximized by changing other parameter dependence of the Hamiltonian, such as changing the direction of the magnetic field. This process in which other parameters of the Hamiltonian change is called the anisotropic magnetocaloric effect  \cite{von}. Thus, the quantity of interest that characterizes MCE is the isothermal entropy; it is expedient to compute this quantity. Since our system has $(2s+1)^2$ discrete energy levels and energy states, the entropy per mole associated with each degree of freedom at infinite temperature $T\to \infty$ is given by
\bea
S_m= 2R\ln (2s+1),
\label{entf}
\eea
where $R$ is the gas constant and $s=9/2$ for the present problem.
\begin{figure}
\centering
\includegraphics[width=3.5in]{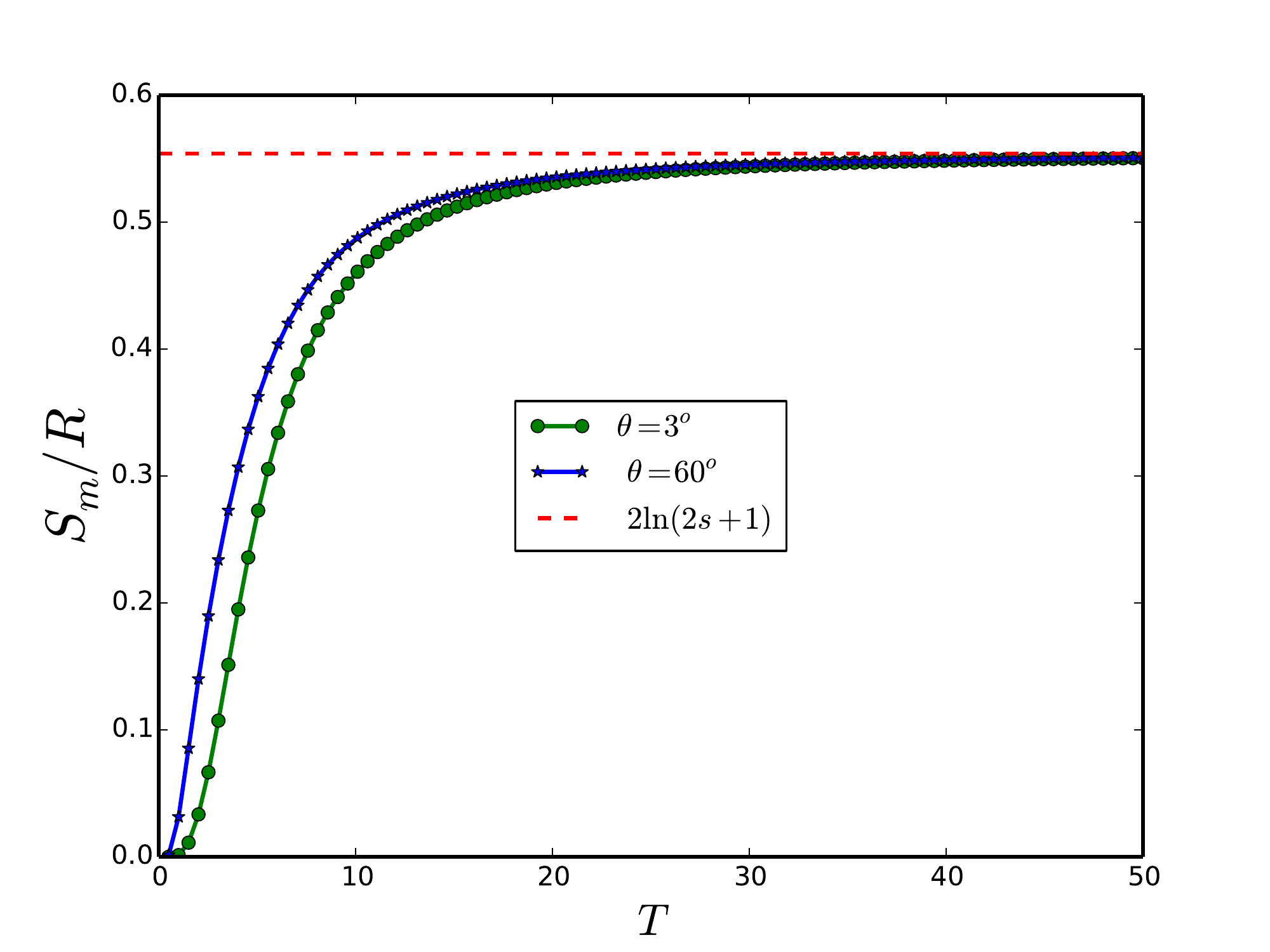}
\caption{Magnetic isothermal entropy, $S_{m}(J/ mol~K)/R$ vs. Temperature, $T([K])$ for $\mathcal{K}=0.77[K],~ J=0.13K, ~ D=0.03[K], ~ \mathcal{J}=3.3\times 10^{-5}[K], ~\mu_B B=1[T]$.}
\label{ent1}
\end{figure}
\begin{figure}
\centering
\includegraphics[width=3.5in]{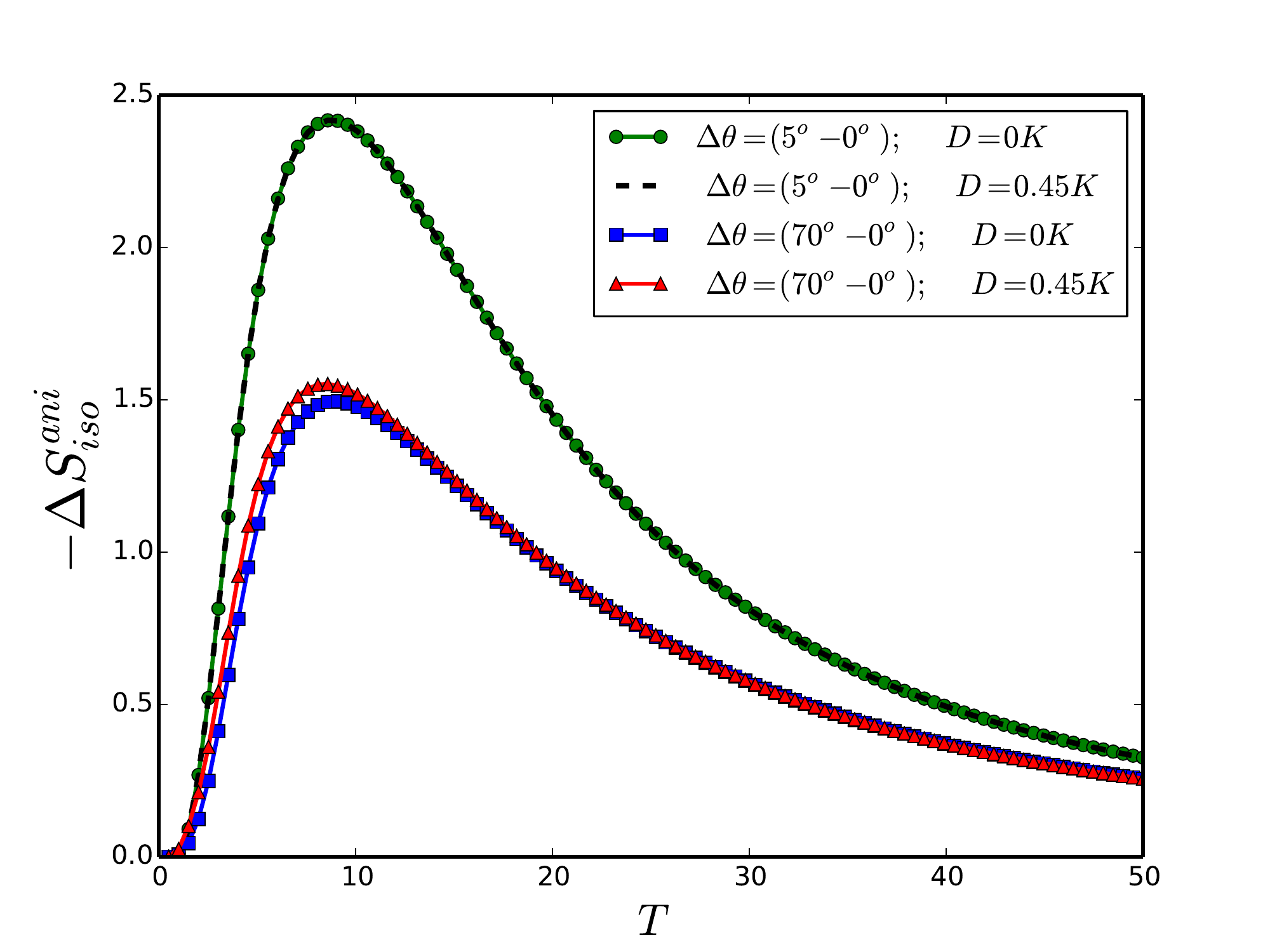}
\caption{Anisotropic magnetic isothermal entropy change, $\Delta S_{iso}(J/ mol~K)/R$ vs. Temperature, $T([K])$ for $\mu_B\Delta B= (5-1)[T]$, ~$\mathcal{K}=0.77[K],~ J=0.13[K], ~\mathcal{J}=3.3\times 10^{-4}[K]$.}
\label{isoth}
\end{figure}
\begin{figure}
\centering
\includegraphics[width=3.5in]{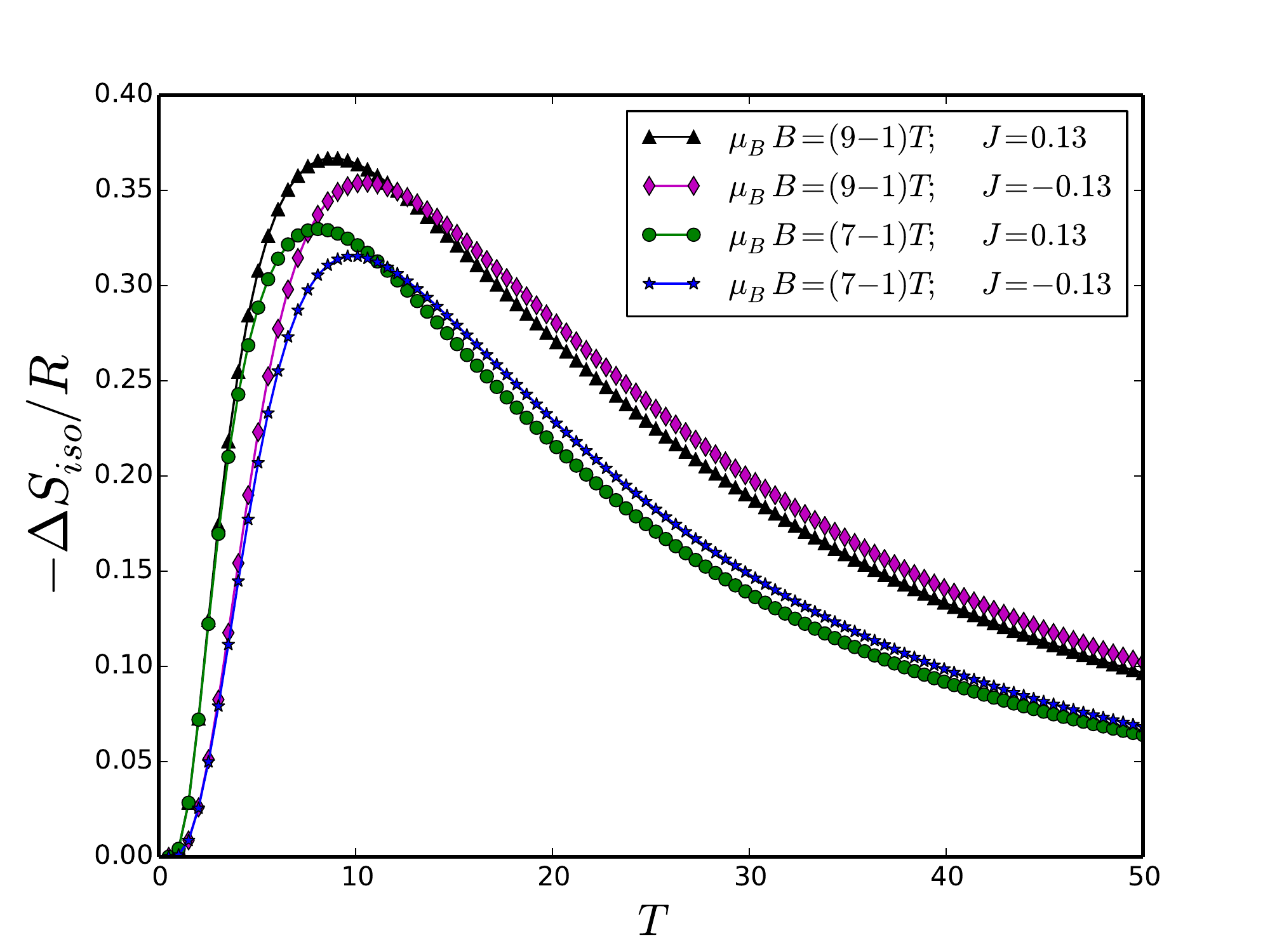}
\caption{Magnetic isothermal entropy change, $\Delta S_{iso}(J/ mol~K)/R$ vs. Temperature, $T([K])$ for $\mathcal{K}=0.77[K],~ D=0.11[K],~  \mathcal{J}=2.3\times 10^{-4}[K],~ \theta=\pi/4$.}
\label{isothe}
\end{figure}
At finite temperature the magnetic contribution to the entropy has the form  \cite{von}:
\bea
S_m(B, T, \theta)= R\lb\ln Z +\frac{\braket{\mathcal{E}}}{T}\rb.
\label{entt}
\eea
Evidently, at infinite temperature $T\to \infty$ the system is highly disordered and all states are equally probable with probability $(2s+1)^{-2}$; thus Eq.\eqref{entt} reduces to Eq.\eqref{entf}. The anisotropic isothermal entropy change $\Delta S_{iso}^{ani}$ follows directly from Eq.\eqref{entt}. As shown in Fig.\eqref{ent1}, the entropy near the easy axis and the hard axis directions at a specific value of the magnetic field approaches the maximum entropy content in Eq.\eqref{entf} as the temperature increases. Increasing the magnetic field causes the spins to orient along the field, thus the entropy decreases.

Fig.\eqref{isoth} shows the anisotropic magnetic isothermal entropy change for several changes in the direction of the magnetic field, with and without the DM interaction. It is evident that the  anisotropic isothermal entropy change is maximized when the magnetic field is close to the easy axis. The influence of a nonzero DM interaction is only visible close to the hard axis direction. An increase in the magnetic field at a fixed angle $\theta$ equally increases the  isothermal entropy as shown in Fig.\eqref{isothe}. In contrast to the demonstration  \cite{marco} that ferromagnetic coupling $J<0$ has a higher isothermal entropy  than the antiferromagnetic coupling $J>0$, we observe in  Fig.\eqref{isothe}, that in the presence of other interactions with a dominant anisotropy,  molecular antiferromagnetic  dimers would be preferred over the ferromagnetic counterparts in modelling MCE, as they possess high isothermal entropy change. 

\subsection{Experimental observations}
The full model we have studied in Eq.\eqref{1}  has not been taken into account experimentally. However, a subset of this model has been used to describe many molecular nanomagnets.  For instance, the sharp peaks in Fig.\eqref{mag} have been observed experimentally \cite{sha} in a related model, using pulsed magnetic fields in the antiferromagnetic dimers [Fe$_2$(salen)$_2$Cl$_2$] and [Fe$_2$(C$_2$O$_4$)(acac)$_4$]   with $s_A=s_B=5/2$. The model Hamiltonian for these dimers, however, contains only the bilinear interaction term in Eq.\eqref{hes}, which evidently is less complicated than  our model in Eq.\eqref{1}.  These peaks were also observed experimentally  \cite{da1} in [Mn$_4$]$_2$ with  $s_A=s_B=9/2$. The model Hamiltonian  for the dimer is devoid of $H_{\text{DM}}$ and $H_{\text{Biquad}}$, with $\mathcal{K}>J$. The parameter values for [Mn$_4$]$_2$ are \cite{da} $\mathcal{K}=0.77K$ and $J=0.13 K$. For experimental purposes, we have used these values in some of our figures.  We believe that the inconsistencies between experimental results and theoretical descriptions could be fixed by taking these higher other terms in consideration. At the level of our theoretical study, the higher order interactions terms seem to have a great influence on the molecules. However,  it  would be  fascinating to deduce experimentally the effects of these additional interactions in physical molecules. 

\section{Conclusion}
In conclusion, we have investigated a molecular dimer model that possesses antisymmetric and biquadractic exchange interaction terms in addition to the customary bilinear Heisenberg interaction; we explicitly demonstrated the influence of these interactions on the quantum behaviour of the system. We observed interesting exotic changes in the quantum behaviour of different phenomena,  which include oscillations of the tunneling splitting, magnetization steps, Schottky anomaly, torque oscillations,  and MCE. Precisely, we found that the DM interaction has a great influence on the magnetization steps and the peaks in the susceptibility, as well as other thermodynamic quantities which are of experimental interest. The biquadractic exchange interaction mainly accounts for some observed deviations in experiments. These interactions should be taken into account in experimental setups as they would provide an accurate description of the system.    It should be of considerable interest to  experimentally investigate how these interaction can provide an exotic behaviour in molecular dimers.


\end{document}